\documentclass[lettersize,journal]{IEEEtran}
\usepackage{amsmath,amsfonts}
\usepackage{algorithmic}
\usepackage{algorithm}
\usepackage{array}
\usepackage[caption=false,font=normalsize,labelfont=sf,textfont=sf]{subfig}
\usepackage{textcomp}
\usepackage{stfloats}
\usepackage{url}
\usepackage{verbatim}
\usepackage{graphicx}
\usepackage{cite}
\usepackage{paracol}
\usepackage{booktabs}
\usepackage{tabularx}
\usepackage{multirow}
\usepackage{xcolor}
\usepackage{ulem}

\begin{document}

\title{Modeling and Resource Optimization for Quantum Oracles}

\author{Zhihang Li, Bo Zhao, Chuanbing Han, Jie Zhao, Jinchen Xu, Guoqiang Shu, Yimin Gao, Woji He and Zheng Shan
\thanks{Zhihang Li, Bo Zhao, Chuanbing Han, Jinchen Xu, Guoqiang Shu, Yimin Gao, Woji He and Zheng Shan are with the Laboratory for Advanced Computing and Intelligence Engineering, Zhengzhou 450001, China (e-mail: shanzhengzz@163.com).}
\thanks{Jie Zhao is with College of Computer Science and Electronic Engineering, Hunan University, Changsha 410082, China. His work was supported by the National Natural Science Foundation of China under Grant No. T2422007 and U24A20235.}}

\markboth{IEEE Transactions on Computers}%
{Li \MakeLowercase{\textit{et al.}}: Modeling and Resource Optimization for Quantum Oracles}


\maketitle

\begin{abstract}
Quantum oracles are fundamental building blocks of many quantum algorithms, and their resource consumption directly affects performance, yet structured description and complexity analysis for their composition are still lacking. In this paper, we introduce the Framework for Oracle Recursion Modeling (FORM), a unified formal abstraction of multi-function composition in quantum oracles: it provides a structured description of the composition layer, makes its gate complexity exactly computable, and turns oracle design into an optimizable tree-construction problem. Based on this model, we propose the ShallowGrow algorithm, which constructs an oracle structure under a given ancilla budget and provably minimizes the number of function evaluations. On Boolean quadratic equation systems, ShallowGrow reduces Qiskit-measured circuit depth by 54.1\% on average relative to the state-of-the-art W-cycle construction, with consistent reductions on the EPFL and ISCAS85 combinational logic networks under scarce ancilla budgets. Furthermore, pebbling-based syntheses trade space against time within the logic network of a function; ShallowGrow extends this trade-off across functions, and integrated with their published circuits it reduces the ancillary qubits of a complete oracle from one per constraint function to logarithmically many. With half as many ancillas as constraint functions, the T-count falls by a factor of 7.9 to 32 relative to the W-cycle-based construction.
\end{abstract}


\section{Introduction}
In the past decades, many quantum algorithms have been proposed for solving different problems, such as machine learning \cite{HOUSSEIN2022116512},\cite{cerezo2022challenges},\cite{LU2024100736}, cryptography \cite{PhysRevLett.95.200502},\cite{Upadhay2019},\cite{Kumar2021StateoftheArtSO}, and finance \cite{gomez2022survey}\cite{naik2025portfolio}. The oracle is a key component of many algorithms, such as Shor's algorithm \cite{shor1996}, Grover's algorithm \cite{grover1996}, and Simon's algorithm \cite{simon1994}. Oracle construction methods have a significant impact on algorithm performance.

There is a significant body of research on oracles, which have been extensively studied as independent and modular components in quantum algorithm design. Oracle techniques have been applied to Boolean equations \cite{liu2022optimizing},\cite{li2024resource}, SAT problems \cite{varmantchaonala2023quantum},\cite{yang2023efficient},\cite{al2024concept}, graph coloring \cite{morris2021simple},\cite{gaspers2023quantum} and AES cryptanalysis \cite{chung2021alternative},\cite{mandal2024implementing}. 
However, these approaches mainly focus on the encoding of specific functions for particular applications.
Other studies consider oracles with a single function \cite{seidel2023automatic},\cite{gaitan2024circuit},\cite{nagy2024novel}, but they face challenges when handling the multiple combinatorial functions common in practice.
In hierarchical reversible logic synthesis, the LHRS framework \cite{soeken2019lut} takes the classical logic network of a single Boolean function as input, partitions it into look-up tables (LUTs), and stores intermediate LUT results on ancilla qubits; the order in which these results are computed and uncomputed is governed by the reversible pebble game \cite{Bennett1989Time}.
Zhang et al.\mbox{\cite{Zhang2025A}} recently proposed a divide-and-conquer pebbling strategy that reduces the total qubit count by up to 90\% compared with the commonly used eager clean-up strategy, at the cost of three to four times more quantum operations.
These methods decompose and schedule the computation of a single function. In contrast, this paper regards each function as an atomic unit and optimizes the hierarchical structure in which the evaluation results of multiple functions are composed under a fixed ancilla budget. Moreover, since such synthesis flows operate on the classical logic network of the target function, they are not applicable when the constituent functions are available only as quantum circuits.

In general, oracle construction involves two related but distinct problems: the encoding of individual functions and the composition of multiple function evaluation results.
The qubit multiplexing technique employs multiple qubits to evaluate different functions and then merges the results to satisfy them simultaneously. Such multiplexing-based composition has been widely adopted in oracle constructions across prior works \cite{liu2022optimizing, li2024resource, gaitan2024circuit, Ambainis2007Improved}. Among existing studies, Li et al.\cite{li2024resource} provide a systematic treatment of the multiplexing-based oracle structure and describe it as a W-cycle, which reduces resource overhead by balancing ancilla count and circuit depth. 
Although this approach provides several preset construction schemes, it mainly relies on discrete structural choices and may lead to suboptimal results when the problem parameters do not match these predefined configurations.
In contrast to single-function synthesis, where trade-offs between total qubit count and gate count have been studied \cite{soeken2019lut},\cite{Zhang2025A}, no comparable adaptive mechanism exists at the composition layer.
Therefore, a systematic framework for modeling the recursive structure of oracle computation and analyzing its quantum circuit complexity is essential.

To address these issues, this paper introduces a Framework for Oracle Recursion Modeling (FORM). Recursive multi-function oracle synthesis has previously lacked a unified modeling foundation, with prior work such as the W-cycle providing only point constructions. FORM is the first to formalize this problem class as an analyzable object: any such construction---existing or future---appears as an instance of a FORM tree and can be measured and compared under a single complexity theory. ShallowGrow is then naturally derived as an adaptive construction within this framework, dynamically balancing ancilla count and circuit depth under a given ancilla budget.
We theoretically prove that, under a given ancilla budget, the ShallowGrow algorithm minimizes the number of function evaluations.
Experiments show average circuit-depth reductions over the state-of-the-art W-cycle of 54.1\% on Boolean quadratic equation systems (Qiskit-measured), and of 43.7\% on the combinational logic networks of the EPFL and ISCAS85 suites under scarce ancilla budgets (analytic). Instantiating the leaves with the published circuits of pebbling-based syntheses yields complete oracles on logarithmically many ancillary qubits rather than one per constraint function; with half as many ancillas as constraint functions, the T-count falls by a factor of 7.9 to 32 relative to the W-cycle-based construction.

\section{Preliminaries: Qubit Multiplexing Technique}
Qubit multiplexing allows a limited number of ancilla qubits to evaluate numerous functions. This technique has been adopted in oracle constructions in many previous studies\cite{liu2022optimizing, li2024resource, gaitan2024circuit, Ambainis2007Improved}.
Each function is first computed and its result is stored in an ancilla qubit. An MCX gate then merges the results of all ancilla qubits into a single target qubit, indicating if all functions are satisfied. Finally, the ancilla qubits are uncomputed and returned to their original states for reuse, while the target qubit records whether the input satisfies all functions.
\begin{figure}[h]
	\centering
	\begin{minipage}[b]{0.24\textwidth}
		\centering
		\includegraphics[width=\textwidth]{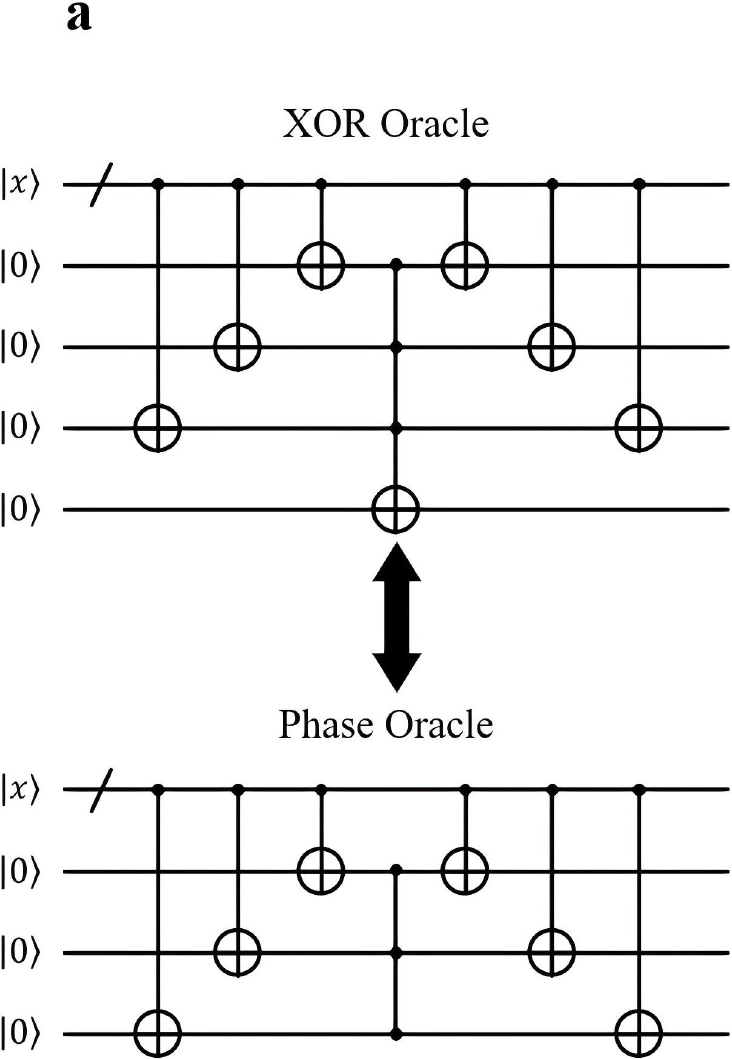}
		\label{fig1a}
	\end{minipage}%
	\hfill
	\begin{minipage}[b]{0.24\textwidth}
		\centering
		\includegraphics[width=\textwidth]{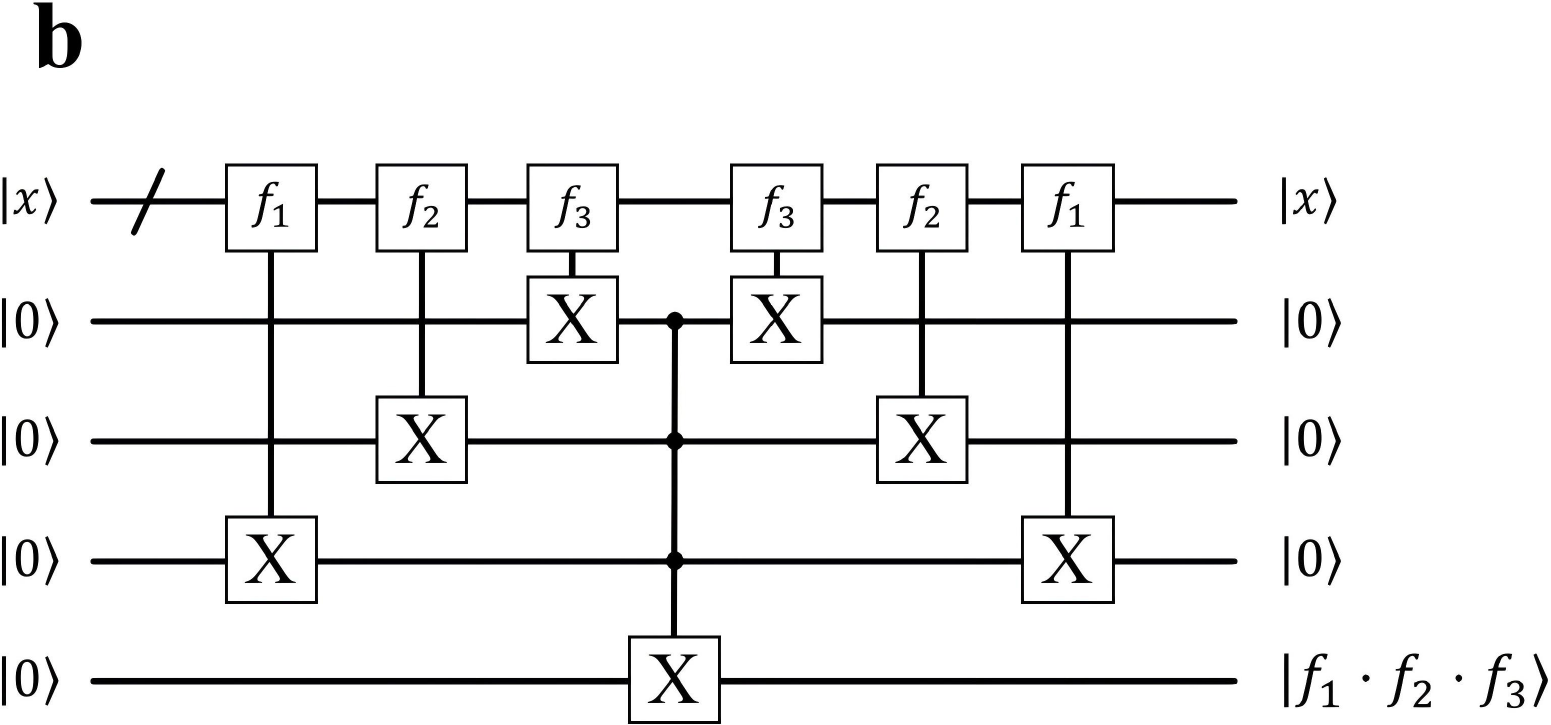}
		\vspace{0.5em}
		
		\includegraphics[width=\textwidth]{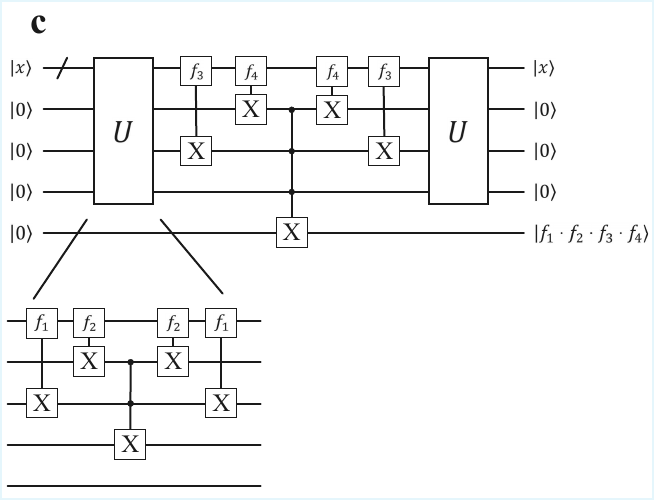}
		\label{fig1bc}
	\end{minipage}
	
	\caption{Framework for constructing quantum oracles using the qubit multiplexing technique.
		(a) Conversion between XOR and phase oracles using multi-controlled X (MCX) and Z (MCZ) gates.
		(b) Evaluation of three constraint functions via qubit multiplexing. Each function is first encoded into an ancilla qubit. The results are then merged onto a target qubit using an MCX gate, after which the ancilla qubits are reset for reuse.
		(c) Hierarchical composition of single-function and multi-function oracles, where smaller oracles are recursively combined into higher-level structures, thereby reducing the required number of ancilla qubits.}
	\label{fig1}
\end{figure}

The quantum oracles studied in this paper implement Boolean functions of the form $f : \{0,1\}^n \to \{0,1\}$. Such oracles are commonly represented in two forms: shift oracles and phase oracles. Without loss of generality, we use the XOR oracle as the primary representation. The XOR oracle is a special case of the shift oracle, which simplifies the presentation and avoids unnecessary technical details. The modeling framework and algorithms proposed in this paper are equally applicable to general shift oracles.

As illustrated in Fig.~\ref{fig1}(a), the XOR oracle and the phase oracle can be converted into each other using multi-controlled X (MCX) and multi-controlled Z (MCZ) gates. More generally, shift oracles and phase oracles can be transformed into each other using the quantum Fourier transform\mbox{\cite{Zhandry2015Quantum}, \cite{Te2025One}}, which further allows the proposed methods to be applied in the phase oracles.

Figure~\ref{fig1}(b) shows an example in which three functions ($m=3$) are evaluated using qubit multiplexing. Importantly, oracles constructed in this manner can themselves serve as modular building blocks. These blocks can be recursively combined to form higher-level oracles, yielding a hierarchical structure that substantially reduces the number of required ancilla qubits, as illustrated in Fig.~\ref{fig1}(c). For clarity, an oracle that computes a single function is called a single-function oracle, while an oracle that computes multiple functions using qubit multiplexing is called a multi-function oracle. These functions are called constraint functions: the multi-function oracle computes their conjunction, so each of them acts as one constraint on the input. For a system of Boolean equations, each equation contributes one constraint function.

In this paper, we use an abstract representation of the constraint functions to develop a general methodology. Our focus is not on how to encode a specific problem function, but on how the computations of multiple functions should be organized and composed once their quantum circuit implementations are given. In practice, users only need to provide the quantum circuit implementation of each function, which can then be incorporated into the proposed framework to construct and analyze the corresponding oracle structure.

\section{The FORM Model}
Constructing an optimal oracle structure requires tools for quantum gate complexity analysis. The recursion depth of an oracle is a key factor affecting its complexity. To address this, we introduce a formal model for oracle structures, termed the Framework for Oracle Recursion Modeling (FORM).

To provide an intuitive understanding, we briefly outline the modeling philosophy of the FORM model. The FORM model aims to provide a structured analytical framework for the complex architecture of multi-function oracles. The key observation is that the composition of functions within a multi-function oracle typically exhibits a recursive structure, which can be naturally represented using a recursive tree.
Specifically, each computation unit is modeled as a node in the tree, while the containment relationships between computation units are represented as edges.
To enable complexity analysis and optimization, each node is associated with several attributes, such as its node depth and the number of child nodes.
Under this modeling approach, even if an oracle appears highly complex at the quantum circuit level, it can be represented by progressively constructing the corresponding FORM tree: each computation unit forms a node, and edges are introduced according to the relationships among these computations.
By traversing and analyzing this tree structure, the quantum gate complexity of the oracle can be systematically evaluated and its structure can be further optimized.

In Section~\ref{sec3.1}, the FORM model and its validity constraints are defined. Section~\ref{sec3.2} then demonstrates how the model can be used to analyze the quantum gate count of an oracle. The FORM model thus provides a systematic foundation, identifies the key factors affecting gate complexity, and establishes a theoretical basis for further optimization.
\subsection{Model Definition}\label{sec3.1}
The FORM model is defined as a triple $T=(V,E,A)$, where $V$ is the set of nodes, each corresponding to a computation unit in the oracle; $E$ is the set of edges, representing the hierarchical containment relationships among computation units; and $A$ is the set of attributes, which assigns to each node a quintuple of properties:

\begin{equation}
	A(v_i)=(s(v_i),d(v_i),\kappa(v_i),c(v_i),\ell(v_i)).
\end{equation}

Definition 1 (Recursive Computation Unit $U_i^{(d)}$):
Let $U_i^{(d)}$ denote the $i$-th computation unit at recursion depth $d$ in the qubit multiplexing technique. When the recursion depth $d$ is not a critical variable, it can be abbreviated as $U_i$.

This model establishes a precise one-to-one correspondence between an oracle's physical structure and its mathematical representation, as defined by the following mappings and constraints:

1. Structural Feature Mapping

(1) Node Correspondence: Each computation unit $U_i^{(d)}$ at recursion depth $d$ corresponds to a node $v_i\in V$ in the tree $T$.

(2) Edge Correspondence: The hierarchical inclusion $U_j^{(d+1)}\subset U_i^{(d)}$ between computation units is represented by an edge $\left(v_i,v_j\right)\in E$. If $\left(v_i,v_j\right)\in E$, node $v_j$ is the child of node $v_i$, and the corresponding computation unit $U_j^{(d+1)}$ is a child computation unit of $U_i^{(d)}$.

2. Attribute Feature Mapping: The attributes of a model node correspond one-to-one with the quantitative features of its associated computation unit:

(1) Node Size: The size $s(v_i)$ of node $v_i$ is the number of ancilla qubits allocated to the computation unit $U_i$, satisfying $s(U_i)=s(v_i)$. The root receives the full ancilla budget, so $s(v_0)=k$.

(2) Node Depth: The depth $d(v_i)$ of node $v_i$, defined as the path length from the root node $v_0$ to $v_i$, corresponds to the recursion depth $d$ of $U_i^{(d)}$, satisfying $d(U_i)=d(v_i)$.

(3) Node Complexity: The complexity $c(v_i)$ of node $v_i$ corresponds to the number of elementary gates in the quantum circuit for the computation unit $U_i$, satisfying $c(U_i)=c(v_i)$.

(4) Covered Leaf Count: The covered leaf count $\ell(v_i)$ of node $v_i$ (i.e., the total number of leaf nodes in the subtree rooted at $v_i)$ equals the total number of underlying constraint functions implemented by $U_i$, satisfying $\ell(U_i)=\ell(v_i)$.

(5) Node Out-degree: The out-degree $\kappa(v_i)$ of node $v_i$ (i.e., the number of direct child nodes) indicates the number of child computation units contained within the computation unit $U_i$, satisfying $\kappa(U_i)=\kappa(v_i)$.

3. Validity Constraints: The correspondence $U_i^{(d)}$ between the FORM model structure and the target oracle requires that the tree $T$ satisfy the following constraints. These constraints ensure that the FORM tree can be consistently mapped to a valid quantum circuit implementation and that ancilla qubit resources can be allocated without conflict during function evaluation.

(1) Monotonicity of Node Size: For any edge $\left(v_i,v_j\right)\in E$, the sizes of the children must be non-negative and must not exceed the size of their parent:
\begin{equation}
	0\leq s(v_j)\leq s(v_i).
	\label{formula2}
\end{equation}

(2) Node Size Distinction among Siblings: If a node $v_i$ has multiple children
$\{v_{j_1},v_{j_2},...\}$, their sizes must be distinct:
\begin{equation}
	s(v_{j_p})\neq s(v_{j_q})\quad(p\neq q).
	\label{formula3}
\end{equation}
This constraint reflects optimal ancilla reuse: sibling nodes are ordered by their output ancilla and thus naturally take distinct sizes; equal sizes would indicate that an earlier sibling failed to exhaust the available intermediate registers.

(3) Leaf Node Constraint: If a node $v_i$ has size 1 or 2, it must be a leaf node, i.e., it has no children:
\begin{equation}
	s(v_i)=1\vee s(v_i)=2\Longrightarrow\kappa(v_i)=0.
	\label{formula4}
\end{equation}

\begin{figure}[h]
	\centering
	\includegraphics[width=0.45\textwidth]{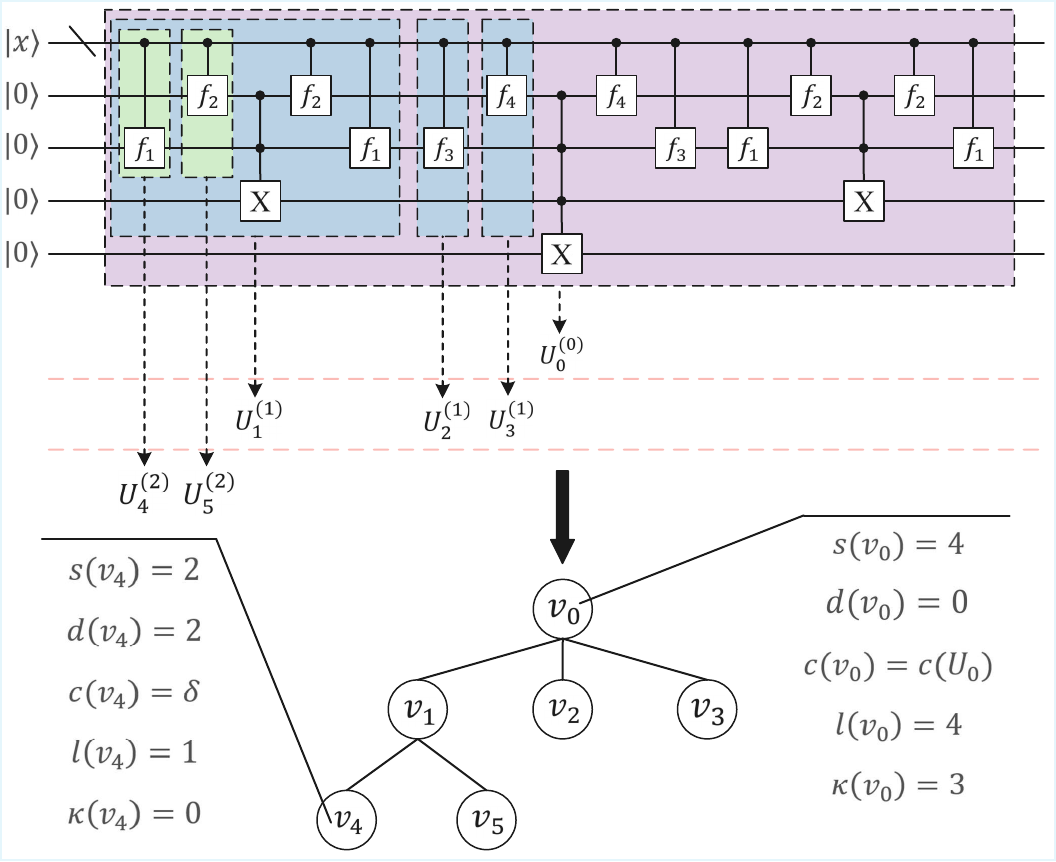}
	\caption{An illustrative example of the FORM model representing an oracle structure with 4 constraint functions and 4 ancilla qubits ($m=4$, $k=4$).}
	\label{FORM_example}
\end{figure}

The FORM model constructs a complete description of the oracle using structural feature mapping, attribute feature mapping, and validity constraints, thereby providing a formal foundation for subsequent quantitative gate complexity analysis and global optimization.  To illustrate the model's operation in practice, an example is provided in Fig.~\ref{FORM_example}.

\subsection{Complexity Analysis}\label{sec3.2}
Let $\delta$ denote the evaluation cost of a single-function oracle, the number of elementary gates in one evaluation; its value depends on the chosen encoding method. The cost of decomposing the MCX gate at a non-leaf node $v_i$ into elementary gates is denoted as $\Gamma(v_i)$, satisfying $\Gamma(U_i)=\Gamma(v_i)$ for the corresponding computation unit $U_i$. This term remains symbolic in subsequent calculations because the proposed model and algorithm are designed to accommodate multiple encoding methods. The value of $\Gamma(v_i)$ depends on the number of control qubits in the MCX gate at node $v_i$, the number of ancilla qubits available for its decomposition, and the specific gate synthesis algorithm employed. Numerous techniques for MCX gate decomposition have been proposed. Without ancilla qubits, an MCX gate can be synthesized with $O(n^2)$ complexity using generalized Peres gates \cite{szyprowski2013low}. With a single ancilla qubit, phase modulation enables a linear resource cost $O(n)$ implementation \cite{he2017decompositions}. Ji Chu et al. \cite{chu2023scalable} achieved linear complexity for MCZ gates; by adding X gates on the target qubit for phase conversion, an equivalent linear-complexity MCX implementation can be derived. Nie et al.\cite{nie2024quantum} further demonstrated that with one ancilla qubit, an optimal decomposition achieving circuit depth and width of $O(n)$ is possible.

For a leaf node $v_i$ (satisfying $\kappa(v_i)=0)$, the complexity is $c(v_i)=\delta$. According to the construction rules of the FORM model, the complexity $c(v_i)$ of a non-leaf node $v_i (\kappa(v_i)\neq0)$ can be recursively calculated from the complexities of all its child nodes and the cost $\Gamma(v_i)$ of the MCX gate used to merge the function results:
\begin{equation}
c(v_i) = \sum_{(v_i,v_j)\in E} 2 \cdot c(v_j) + \Gamma(v_i), \quad (\kappa(v_i) \neq 0). 	\label{eq:FORM_complexity}
\end{equation}

The total gate count of the entire oracle structure $O$ (i.e., the complexity of the root node $v_0$) can be obtained by traversing all computation units and summing their individual complexities.

Theorem~1 (FORM Global Complexity). Let $\mathcal{U}_{\mathrm{base}} := \{U_{f_i}\}$ denote the set of all single-function oracles, and $\mathcal{U}_{\mathrm{comp}} := \{U_{F_i}\}$ denote the set of all multi-function oracles. The global complexity $c(O)$ is expressed as:
\begin{equation}
	c(O) = \sum_{U_{i} \in ~\mathcal{U}_{\mathrm{base}}} 2^{d(U_{i})} \cdot \delta + \sum_{U_{i} \in ~\mathcal{U}_{\mathrm{comp}}} 2^{d(U_{i})} \cdot \Gamma(U_{i}).
\end{equation}

In the FORM model, this is equivalent to the complexity of the root node:
\begin{equation}
	c(v_0)=\sum_{v\in V,\kappa(v)=0}2^{d(v)}\cdot\delta+\sum_{u\in V,\kappa(u)\neq0}2^{d(u)}\cdot\Gamma(u).
	\label{formula7}
\end{equation}

Nodes are classified as leaf nodes when $\kappa(v) = 0$, representing single-function oracles, and as non-leaf nodes when $\kappa(u) \neq 0$, representing multi-function oracles. The inductive proof of this theorem is as follows:

\textbf{Base case.} 
For a single-leaf tree with root $v_0$ where $\kappa(v_0)=0$, we have $s(v_0)=s(U_0)$ (number of ancilla qubits), $d(v_0)=0$, $c(v_0)=\delta$ and $\ell(v_0)=1$.

Substituting into Equation \ref{formula7} yields:
\begin{equation}
	\sum_{\kappa(v)=0}2^{d(v)}\delta=2^0\delta=\delta,\quad 
	\sum_{\kappa(u)\neq0}2^{d(u)}\Gamma(u)=0.
\end{equation}

Hence $c(v_0)=\delta$, which is consistent with the definition.

\medskip
\textbf{Inductive hypothesis.} 
Assume Theorem~1 holds for any FORM tree of height $\leq h$.

\medskip
\textbf{Inductive step.} 
Consider a tree $T$ of height $h+1$ with root $v_0$ (where $(\kappa(v_0)\neq0)$).
For each child $v_j$ with subtree $T_j=(V_j,E_j,A_j)$ of height $\leq h$:
\begin{itemize}
	\item If $\kappa(v_j)=0$, then $c(v_j)=\delta$.  
	\item If $\kappa(v_j)\neq0$, by induction,
	\begin{equation}
		c(v_j)=\sum_{\substack{v\in V_j\\ \kappa(v)=0}}2^{d(v|T_j)}\cdot\delta
		+\sum_{\substack{u\in V_j\\ \kappa(u)\neq0}}2^{d(u|T_j)}\cdot\Gamma(u),
	\end{equation}
\end{itemize}
where $d(v|T_j)$ denotes the node depth of $v$ within subtree $T_j$, and $d(v|T_j)=d(v)-1$ (similarly for $u$).

Substituting these expressions for the child nodes into the complexity formula for $c(v_0)$ yields:
\begin{equation}
	\begin{aligned}
		c(v_0) &= \sum_{j=1}^{\kappa(v_0)} 2 \cdot \!\left(\sum_{\substack{v \in V_j \\ \kappa(v)=0}} 2^{d(v)-1} \cdot \delta
		+ \sum_{\substack{u \in V_j \\ \kappa(u) \neq 0}} 2^{d(u)-1} \cdot \Gamma(u) \right) \\
		&\quad + \Gamma(v_0).
	\end{aligned}
\end{equation}

Simplifying,
\begin{equation}
	\begin{aligned}
		c(v_0) &= \sum_{j=1}^{\kappa(v_0)} \biggl( \sum_{\substack{v \in V_j \\ \kappa(v) = 0}} 2^{d(v)}\cdot\delta
		+ \sum_{\substack{u \in V_j \\ \kappa(u) \neq 0}} 2^{d(u)}\cdot\Gamma(u) \biggr) \\
		&\quad + \Gamma(v_0) \cdot 2^{d(v_0)}.
	\end{aligned}
\end{equation}

Thus,
\begin{equation}
	c(v_0)=\sum_{\kappa(v)=0}2^{d(v)}\delta
	+\sum_{\kappa(u)\neq0}2^{d(u)}\Gamma(u),
\end{equation}
which is consistent with Equation \ref{formula7}.

\begin{figure}[h]
	\centering
	\includegraphics[width=0.45\textwidth]{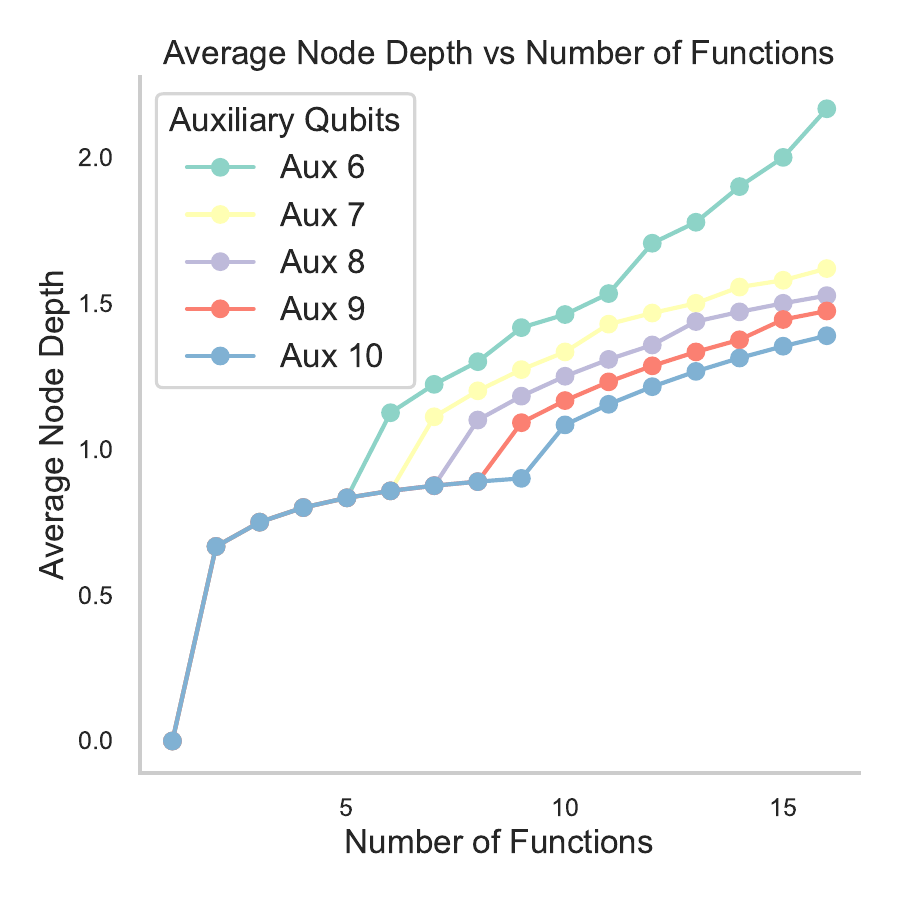}
	\caption{Average node depth versus the number of constraint functions for different ancilla qubit counts. The curves represent results for ancilla qubit counts $k$ ranging from 6 to 10.}
	\label{fig2}
\end{figure}

In this expression, the node depth $d$ has an exponential effect on the number of quantum gates, making it a critical factor for optimization. In the ShallowGrow algorithm introduced in Section~\ref{secShallowGrow}, the average value of $d$ shows an approximately logarithmic relationship with the number of constraint functions. As the number of functions increases, the growth of the average node depth slows, exhibiting a sub-linear trend (see Fig.~\ref{fig2}). For problem instances with up to 15 constraint functions, the average node depth remains below 2, resulting in a minimal impact on the overall cost. This property is key to the excellent performance of the ShallowGrow algorithm.

\section{ShallowGrow Algorithm}\label{secShallowGrow}
Building upon the FORM model and its complexity framework, we introduce the ShallowGrow algorithm, which incrementally generates FORM tree structures to minimize root-node complexity under a given ancilla budget. Section~\ref{sec4.1} details the ShallowGrow algorithm's procedure, and Section~\ref{sec4.2} proves that it minimizes the number of function evaluations.

\subsection{Algorithm Process}\label{sec4.1}
In the ShallowGrow algorithm, the FORM tree $T_k(m)$ is constructed incrementally by progressively adding function nodes. To efficiently select the node for expansion, the algorithm employs two prioritization strategies. Strategy 1 (the minimum-depth strategy) selects the node with the smallest node depth, thereby controlling the growth scale of the tree. When multiple nodes share the same node depth, Strategy 2 (the maximum-size strategy) is applied, selecting the node with the largest size to maximize its capacity for accommodating child nodes. The rationale for these strategies stems from the complexity formula in Equation \ref{eq:FORM_complexity}. Strategy 1 minimizes node depth. Because expanding a leaf node (i.e., splitting it) incurs the cost $\Gamma$ of one additional MCX gate, Strategy 2 minimizes the total number of such expansions required for a given number of constraint functions. This results in fewer non-leaf nodes contributing to the sum in Equation \ref{eq:FORM_complexity}, thereby reducing the total cost. The combined use of these strategies guarantees the optimality of the generated structure $T_k(m)$; a rigorous proof is provided in Section~\ref{sec4.2}. Upon completion of the tree, a single post-order traversal computes the complexity $c(v_i)$ and the number of leaf nodes $\ell(v_i)$ for each node $v_i$.

\begin{algorithm}[H]
	\caption{ShallowGrow Algorithm (Part 1).}
	\label{algo1}
	\begin{algorithmic}
		\STATE \textbf{Input:} $m$: number of constraint functions, $k$: ancilla budget
		\STATE \textbf{Output:} Tree $T = (V, E, A)$ with attributes $s(v), d(v), \kappa(v), c(v), \ell(v)$
		\STATE Initialize $T = (V, E, A)$
		\STATE $V \gets \{v_0\}$; $E \gets \emptyset$
		\STATE $A(v_0) \gets (s = k, d = 0, \kappa = 0, c = \text{null}, \ell = \text{null})$
		\STATE $m_{\text{cur}} \gets 1$
		\STATE \textbf{while} $m_{\text{cur}} < m$ \textbf{do}
		\STATE \hspace{0.5cm} \textbf{if} tree $T$ is saturated \textbf{then}
		\STATE \hspace{1cm} $C \gets$ findCandidateLeaves$(T)$
		\STATE \hspace{1cm} \textbf{if} $C \neq \emptyset$ \textbf{then}
		\STATE \hspace{1.5cm} $C' \gets \{v \in C \mid d(v)=\min_{u\in C} d(u)\}$
		\STATE \hspace{1.5cm} select $v_i \in C'$ with max $s(v_i)$
		\STATE \hspace{1cm} Create $v_p$ with 
		\STATE \hspace{1.5cm} $s = s(v_i)-1$, $d = d(v_i)+1$
		\STATE \hspace{1.5cm} $\kappa = 0$, $c = \text{null}$, $\ell = \text{null}$
		\STATE \hspace{1cm} Create $v_q$ with
		\STATE \hspace{1.5cm} $s = s(v_i)-2$, $d = d(v_i)+1$
		\STATE \hspace{1.5cm} $\kappa = 0$, $c = \text{null}$, $\ell = \text{null}$
		\STATE \hspace{1cm} $V \gets V \cup \{v_p, v_q\}$; 
		\STATE \hspace{1.1cm} $E \gets E \cup \{(v_i, v_p), (v_i, v_q)\}$
		\STATE \hspace{1cm} $\kappa(v_i) \gets \kappa(v_i) + 2$
		\STATE \hspace{0.5cm} \textbf{else}
		\STATE \hspace{1cm} $v_i \gets$ findUnsaturatedNode$(T)$
		\STATE \hspace{1cm} Create $v_j$ with
		\STATE \hspace{1.5cm} $s = s(v_i)-\kappa(v_i)-1$, $d = d(v_i)+1$
		\STATE \hspace{1.5cm} $\kappa = 0$, $c = \text{null}$, $\ell = \text{null}$
		\STATE \hspace{1cm} $V \gets V \cup \{v_j\}$; $E \gets E \cup \{(v_i, v_j)\}$
		\STATE \hspace{1cm} $\kappa(v_i) \gets \kappa(v_i) + 1$
		\STATE \hspace{0.5cm} \textbf{end if}
		\STATE \hspace{0.5cm} $m_{\text{cur}} \gets m_{\text{cur}} + 1$
		\STATE \textbf{end while}
		\STATE PostorderTraversal$(T, v_0)$
	\end{algorithmic}
\end{algorithm}
During construction, nodes are classified as follows. A \textbf{saturated node} is one that cannot accept more child nodes, i.e., $\kappa(v) = s(v)-1$ or $s(v) \le 2$; an \textbf{unsaturated node} is a node that already contains some child nodes but can still accommodate additional ones, i.e., $0 < \kappa(v) < s(v)-1$; a \textbf{candidate node} is any node capable of accepting new child nodes. This condition is equivalent to $s(v) > 2$ and $\kappa(v) < s(v)-1$. Furthermore, a FORM tree $T_k(m)$ is called a \textbf{saturated tree} if all its non-leaf nodes are saturated nodes.

\begin{figure}[h]
	\centering
	\includegraphics[width=0.45\textwidth]{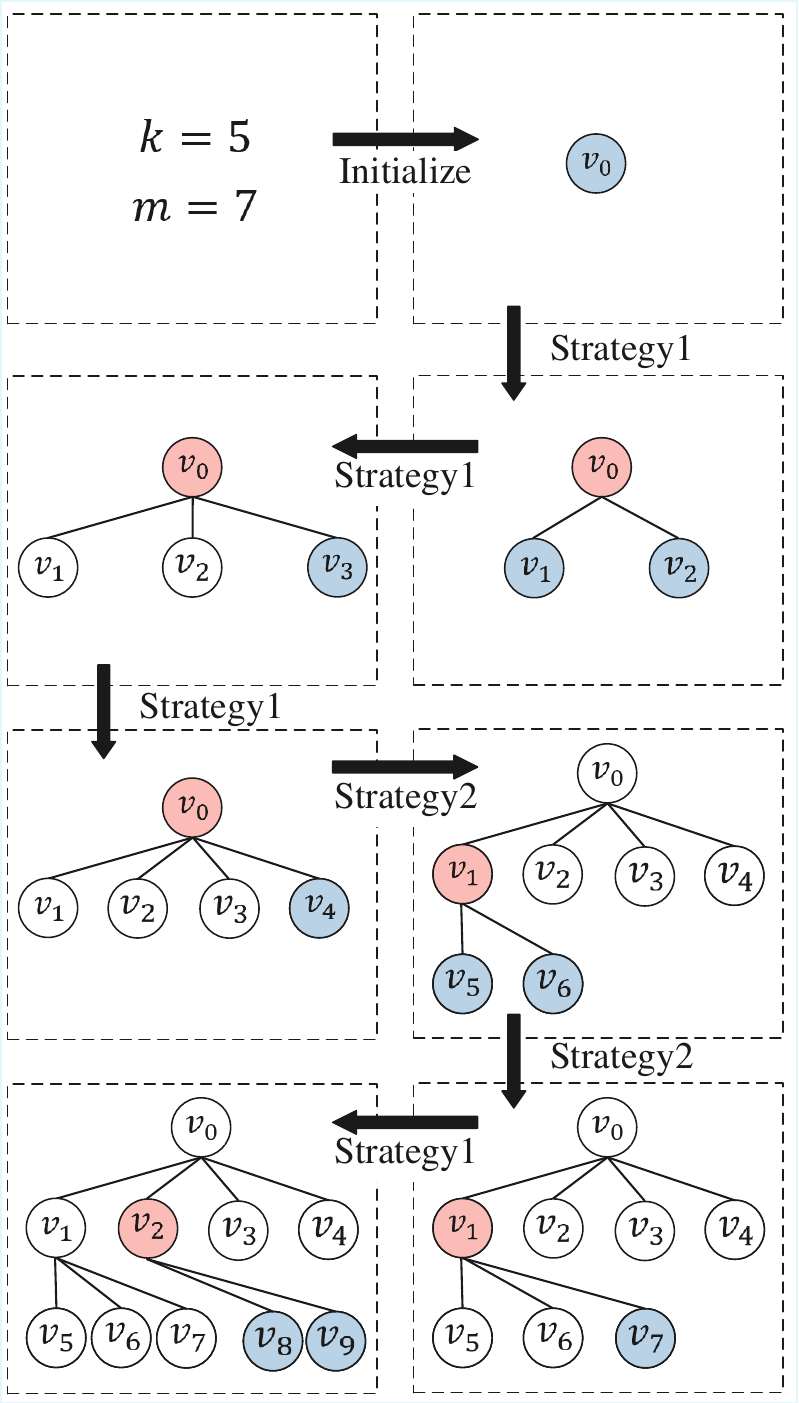}
	\caption{Incremental construction of the FORM tree $T_k(m)$ using the ShallowGrow algorithm, illustrated with a small problem instance. At each step, the parent node selected for expansion by the ShallowGrow algorithm is highlighted in red, and the newly added child node is highlighted in blue. The iterative process continues until the tree contains 7 leaf nodes, corresponding to the 7 constraint functions to be implemented.}
	\label{ShallowGrow_example}
\end{figure}

The pseudocode for the ShallowGrow algorithm is presented in Algorithm \ref{algo1}, which has a time complexity of $O(m)$. A step-by-step example illustrating the algorithm's execution for solving 7 constraint functions with 5 ancilla qubits is provided in Fig.~\ref{ShallowGrow_example}.
\begin{algorithm}[H]
	\ContinuedFloat
	\caption{ShallowGrow Algorithm (Part 2).}
	\begin{algorithmic}
		\STATE \textbf{function} findCandidateLeaves$(T)$
		\STATE \hspace{0.5cm} $C \gets \emptyset$
		\STATE \hspace{0.5cm} \textbf{for each} $v \in V$ \textbf{do}
		\STATE \hspace{1cm} \textbf{if} $\kappa(v)=0$ \textbf{and} $s(v)>2$ \textbf{and} $v$ is \textbf{not} saturated
		\STATE \hspace{1cm} \textbf{then} 
		\STATE \hspace{1.5cm} $C \gets C \cup \{v\}$
		\STATE \hspace{0.5cm} \textbf{end for}
		\STATE \hspace{0.5cm} \textbf{return} $C$
		\STATE
		\STATE \textbf{function} findUnsaturatedNode$(T)$
		\STATE \hspace{0.5cm} \textbf{for each} $v \in V$ \textbf{do}
		\STATE \hspace{1cm} \textbf{if} $s(v) > 2$ \textbf{and} $0<\kappa(v)<s(v)-1$ \textbf{and} $v$ is \textbf{not} saturated
		\STATE \hspace{1cm} \textbf{then return} $v$
		\STATE \hspace{0.5cm} \textbf{end for}
		\STATE
		\STATE \textbf{function} PostorderTraversal$(T, v)$
		\STATE \hspace{0.5cm} \textbf{if} $\kappa(v) = 0$ \textbf{then} 
		\STATE \hspace{1cm} $c(v) \gets \delta$, $\ell(v) \gets 1$
		\STATE \hspace{0.5cm} \textbf{else}
		\STATE \hspace{1cm} $c_{\text{sum}} \gets 0$, $\ell_{\text{sum}} \gets 0$
		\STATE \hspace{1cm} \textbf{for each child} $v_j$ \textbf{do}
		\STATE \hspace{1.5cm} PostorderTraversal$(T, v_j)$
		\STATE \hspace{1.5cm} $c_{\text{sum}} \gets c_{\text{sum}} + 2 \cdot c(v_j)$
		\STATE \hspace{1.5cm} $\ell_{\text{sum}} \gets \ell_{\text{sum}} + \ell(v_j)$
		\STATE \hspace{1cm} \textbf{end for}
		\STATE \hspace{1cm} $c(v) \gets c_{\text{sum}} + \Gamma(v)$, $\ell(v) \gets \ell_{\text{sum}}$
		\STATE \hspace{0.5cm} \textbf{end if}
	\end{algorithmic}
\end{algorithm}

\subsection{Optimality Proof for Leaf-Node Cost}\label{sec4.2}
In multi-function oracle construction, the computational cost in our model is naturally divided into leaf-node cost and non-leaf-node cost. Leaf nodes represent function evaluations, so the leaf-node cost measures the number of repeated function evaluations, whereas non-leaf nodes correspond to MCX gates used to combine intermediate results. In this proof, we focus on the leaf-node cost term $\sum_{\kappa(v)=0}2^{d(v)}\cdot\delta$; the non-leaf-node cost is analyzed separately in Section~\ref{sec4.3}. The quantity $\sum_{\kappa(v)=0} 2^{d(v)}$ depends only on the tree structure and is independent of the specific evaluation cost of each leaf, so the optimality result below holds for arbitrary leaf costs.

We prove by contradiction that the cost $\sum_{\kappa(v)=0}2^{d(v)}\cdot\delta$ is minimized. Assume, for contradiction, that for the FORM tree $T_k(m)$ constructed by the ShallowGrow algorithm, there exists a valid tree $T_k'(m)$ such that $c(T_k'(m))<c(T_k(m))$. Since $T_k(m)\neq T_k'(m)$, we define

\begin{align}
	X &= \{v \mid v \in V(T_{k}(m)),\, v \notin V(T_{k}^{\prime}(m))\}, \\
	Y &= \{u \mid u \in V(T_{k}^{\prime}(m)),\, u \notin V(T_{k}(m))\}.
\end{align}

By the minimum-depth strategy of the ShallowGrow algorithm, for any $v_a\in X$ and $v_b\in Y$, we have $d(v_a)\leq d(v_b)$. We construct an iteration sequence $\{T^{(t)}\}_{t=0}^N$ with $T^{(0)}=T_k^{\prime}(m)$, such that each adjustment ensures $c(T^{(t)})\geq c(T^{(t-1)})$.

\noindent\textbf{1. Structural Adjustment.}

In each iteration, we choose
\begin{equation}
	v_a=\arg\min_{v\in X}d(v), v_b=\arg\max_{u\in Y}d(u).
\end{equation}

\emph{Case 1.} If $d(v_{a}) < d(v_{b})$ and the parent $v_{a}'$ of $v_{a}$ in $T_{k}(m)$ is a leaf node in $T^{(t)}$, we delete $v_{b}$ and add $v_{a}$ and its sibling $v_{a*}$ as children to $v_{a}'$. Then, we update the sets $X$ and $Y$. Since $d(v_b)>\max\{d(v_a), d(v_{a*})\}$,
\begin{equation}
	2^{d(v_b)}\delta \ge (2^{d(v_a)}+2^{d(v_{a*})})\delta,
\end{equation}
thus the cost $c(T^{(t)})$ does not increase.

\emph{Case 2.} If $v_a'$ is non-leaf in $T^{(t)}$, delete $v_b$, add $v_a$ to $v_a'$, and update $X,Y$. As $d(v_a)<d(v_b)$,
\begin{equation}
	2^{d(v_b)}\delta > 2^{d(v_a)}\delta.
\end{equation}

Hence, the cost $c(T^{(t)})$ does not increase.

\noindent\textbf{2. Terminal State.}

\emph{Case 1.} If $X=Y=\emptyset$, then $T^{(N)}=T_k(m)$ and
\begin{equation}
	c(T_k(m))=c(T^{(N)})\geq c(T^{(0)})=c(T_k^{\prime}(m)),
\end{equation}
contradicting our initial assumption.

\emph{Case 2.} If all remaining nodes in $X,Y$ have node depth $d^*$, then all are leaves (otherwise deeper nodes exist, contradiction). By the maximum-size strategy of the ShallowGrow algorithm at equal node depth, the expansion process minimizes the total MCX cost, which ensures that $c(T_k(m)) \leq c(T^{(t)})$. Thus, we have $c(T_k^{\prime}(m))=c(T^{(0)}) \geq c(T^{(N)})=c(T_k(m))$, which again leads to a contradiction.

In all cases, the initial assumption leads to a contradiction. Therefore, the leaf-node cost of the tree $T_k(m)$ generated by the ShallowGrow algorithm is minimized.

\subsection{Complexity Analysis for Non-Leaf-Node Cost}\label{sec4.3}

This section provides a supplementary complexity analysis of the non-leaf-node cost term $\sum_{\kappa(u)\neq 0}2^{d(u)}\cdot\Gamma(u)$ to better understand the overall complexity behavior. While we do not provide a formal optimality proof for the non-leaf-node cost term, this term is still reduced to some extent under the ShallowGrow algorithm.

The ShallowGrow algorithm minimizes the leaf-node cost term $\sum_{\kappa(v)=0}2^{d(v)}\cdot\delta$, i.e., the exponential sum over the node depths of the leaves. This optimization also benefits the non-leaf-node cost term. Since every non-leaf node lies on a path to a leaf, its node depth cannot exceed that of its descendant leaf nodes. Therefore, reducing leaf-node depths directly constrains the exponential factor $2^{d(u)}$ in the non-leaf-node cost.

For a non-leaf node $u$, $\Gamma(u)$ denotes the decomposition cost of the corresponding MCX gate, which depends on the number of control qubits $\kappa(u)$ (equivalently, the number of child nodes) and grows polynomially, i.e., $\Gamma(u)=\mathcal{O}(\kappa(u)^p)$ for $p\in\{1,2\}$ \cite{szyprowski2013low},\cite{he2017decompositions},\cite{chu2023scalable},\cite{nie2024quantum}. Since the total number of constraint functions is fixed to $m$, the number of functions that can be assigned to any non-leaf node is inherently limited. At the same time, under a fixed ancilla budget, the number of qubits that can be allocated to a non-leaf node is constrained by the available ancillas $k$. Hence, the range of $\kappa(u)$ is jointly bounded by $m$ and $k$. Moreover $\Gamma(u)$ grows only polynomially with $\kappa(u)$, its variation is much smaller than that of the exponential factor $2^{d(u)}$. As a result, the non-leaf-node cost term is governed primarily by node depth.
Thus, minimizing the leaf-node cost $\sum_{\kappa(v)=0}2^{d(v)}\cdot\delta$ also constrains the non-leaf-node cost $\sum_{\kappa(u)\neq 0}2^{d(u)}\cdot\Gamma(u)$.

\section{Experimental Evaluation}\label{sec:exp}
We evaluate the constructed oracles in three settings: circuit depth against the W-cycle construction under equal ancilla budgets, the space--depth trade-off over the full ancilla-budget axis, and the resource consumption (T-count and total qubit count) of complete oracle constructions on published leaf-level syntheses.

\subsection{Experimental Setup}\label{sec:exp-setup}
Two instance families are used. The first covers the ten arithmetic instances of the EPFL combinational logic synthesis benchmarks \cite{amaru2015epfl} (adder, bar, div, hyp, log2, max, multiplier, sin, sqrt, square), which are publicly available and commonly used to evaluate logic synthesis algorithms. The ten ISCAS85 combinational circuits (c432, c499, c880, c1355, c1908, c2670, c3540, c5315, c6288, c7552) are also included. A circuit with $m$ primary outputs is read as a system of $m$ constraint functions, one per output cone; $m$ ranges from 7 (c432) to 140 (c2670), and cone sizes span from zero AND nodes to $2.1\times10^{5}$ AND nodes (hyp). The second family consists of the Boolean quadratic equation systems published with the W-cycle study \cite{li2024resource}, with the number of variables $n\in\{15,20,25\}$ and $m=5$, $7$, $9$ equations per Grover iteration, the scale of the original comparison.

The compared methods fall into two roles. At the composition layer, ShallowGrow is compared with FORM-B and with the W-cycle construction. FORM-B is the balanced instantiation of the FORM model: the functions of every node are divided as evenly as possible among its children---a node holding $f$ functions over $c$ children assigns $\lfloor f/c\rfloor$ or $\lceil f/c\rceil$ of them to each. FORM-B serves as a same-model ablation: it holds the FORM modeling framework fixed and uses only a naive balanced instantiation, thereby isolating ShallowGrow's algorithmic contribution from FORM's modeling contribution. The W-cycle is implemented after Algorithms~1--2 of \cite{li2024resource} for an arbitrary recursion level: the constraint functions fill the W-cycle nodes in construction order and unfilled W-cycle nodes are removed. Levels one to three (W-L1, W-L2, W-L3) cover all budgets of Section~\ref{sec:exp-comp}, and the tightest budgets of Sections~\ref{sec:exp-tradeoff} and \ref{sec:exp-stack} require levels up to W-L7. At the leaf layer, a single constraint function is encoded either by direct XAG (XOR--AND--inverter graph) mapping (one Toffoli per AND node, computed and uncomputed inside the leaf, following the reversible-simulation scheme of Bennett \cite{bennett1973logical}), by LHRS \cite{soeken2019lut}, or by the divide-and-conquer pebbling strategy of Zhang et al.\ \cite{Zhang2025A}. Sections~\ref{sec:exp-comp} and \ref{sec:exp-tradeoff} fix the leaf layer to direct XAG mapping and vary the composition layer only; Section~\ref{sec:exp-stack} varies both layers.

Sections~\ref{sec:exp-comp} and \ref{sec:exp-tradeoff} report analytic gate count and circuit depth, obtained by evaluating the recursion of Theorem~1 under the gate-count and depth weights given below and the MCX decomposition of \cite{barenco1995elementary}. The W-cycle is itself an instance of the FORM model (implemented after Algorithms~1--2 of \cite{li2024resource} at every level), so Theorem~1 evaluates its complexity by the same recursion as ShallowGrow, and the same weights apply throughout. Benchmark oracles reach circuit depths of $10^{9}$ with up to $2.1\times10^{5}$ AND nodes in a single constraint function, beyond full circuit construction. Where the scale permits full construction, the oracles are built in Qiskit and the depth is measured on the decomposed circuits including cross-qubit parallelism (Table~\ref{tab:legacy}).

Gates in $\{H, T, T^{\dagger}, S, S^{\dagger}, X, \mathrm{CNOT}\}$ count one each toward gate count and circuit depth; a Toffoli gate counts 15 gates and contributes 12 to the circuit depth; an MCX with $\kappa\ge3$ controls is decomposed into $4(\kappa-2)$ Toffoli gates using $\kappa-2$ dirty ancillas borrowed from the input register \cite{barenco1995elementary}. The evaluation cost of an output cone with $a$ AND nodes, $x$ XOR nodes, and $i$ inverters is $\delta = 2(15a+x+i)+1$ gates: the cone is computed, copied onto its composition ancilla, and uncomputed. Because the root of an oracle is merged by an MCZ and consumes no result qubit, the root of a size-$k$ tree holds up to $k$ children, while any other non-leaf node of size $s$ reserves its top qubit for the MCX result and holds up to $s-1$ children; this convention is applied identically to every method. For Section~\ref{sec:exp-stack}, each leaf synthesizer is calibrated on each benchmark by its per-node cost $\rho = T_{\mathrm{pub}}/N_{\mathrm{AND}}$, taken from Table~III of \cite{soeken2019lut} (ESOP heuristic def\_wo4, hybrid mapping, minimum-T row) and Table~II of \cite{Zhang2025A} (cut size 6, divide-and-conquer), so a cone with $a_i$ AND nodes costs $\delta_i = \rho\, a_i$ T gates.

Throughout the experiments, below saturation ($k<m$) the ancilla budget is an upper bound: every deterministic construction takes its best ancilla register $k'\le k$ and leaves the remaining ancillas idle. At $k\ge m$ placement is trivial and every construction returns the same flat tree, one function per ancilla. In all structures the same rule assigns the costliest function to the shallowest leaf.
\subsection{Composition-Layer Comparison}\label{sec:exp-comp}
\begin{table*}[!t]
	\centering
	\caption{Circuit depth of the ShallowGrow oracle and its reduction (\%) against FORM-B and the W-cycle (W-cyc.), EPFL circuits in the upper block and ISCAS85 in the lower. FORM-B: the balanced FORM tree, with the functions of every node distributed evenly over its children and no placement optimization. $k$: the ancilla budget; below saturation the budget is an upper bound, and every construction takes its best ancilla register $k'\le k$, leaving the remaining ancillas idle. The three budgets per circuit correspond to the first three W-cycle levels: ${\approx}\sqrt{2m}$ (W-L3), $\lceil m/2\rceil$ (W-L2), and $m$ (W-L1, saturation).}
	\label{tab:comp}
	\footnotesize
	\renewcommand{\arraystretch}{1.15}
	\setlength{\tabcolsep}{6pt}
	\begin{tabular}{lrrrrrrrrrrr}
		\toprule
		 &  & \multicolumn{4}{c}{$k\approx\sqrt{2m}$} & \multicolumn{4}{c}{$k=\lceil m/2\rceil$} & \multicolumn{2}{c}{$k=m$$^{\dagger}$} \\
		\cmidrule(lr){3-6}\cmidrule(lr){7-10}\cmidrule(lr){11-12}
		Circuit & $m$ & $k$ & Depth & FORM-B & W-cyc. & $k$ & Depth & FORM-B & W-cyc. & $k$ & Depth \\
		\midrule
		adder & 129 & 16 & $3.6{\times}10^{6}$ & 6.9 & 46.9 & 65 & $2.3{\times}10^{6}$ & 35.7 & 36.8 & 129 & $1.8{\times}10^{6}$ \\
		bar & 128 & 16 & $6.0{\times}10^{6}$ & 19.1 & 45.5 & 64 & $4.3{\times}10^{6}$ & 23.5 & 24.2 & 128 & $2.8{\times}10^{6}$ \\
		div & 128 & 16 & $2.6{\times}10^{8}$ & 4.5 & 48.1 & 64 & $1.7{\times}10^{8}$ & 35.3 & 36.0 & 128 & $1.3{\times}10^{8}$ \\
		hyp & 128 & 16 & $2.3{\times}10^{9}$ & 17.1 & 45.7 & 64 & $1.6{\times}10^{9}$ & 26.4 & 27.1 & 128 & $1.1{\times}10^{9}$ \\
		log2 & 32 & 8 & $1.1{\times}10^{8}$ & 9.8 & 44.1 & 16 & $7.8{\times}10^{7}$ & 20.3 & 22.9 & 32 & $5.0{\times}10^{7}$ \\
		max & 130 & 16 & $3.6{\times}10^{7}$ & 17.7 & 44.3 & 65 & $2.6{\times}10^{7}$ & 23.6 & 24.2 & 130 & $1.7{\times}10^{7}$ \\
		multiplier & 128 & 16 & $1.8{\times}10^{8}$ & 3.2 & 47.5 & 64 & $1.1{\times}10^{8}$ & 39.6 & 40.6 & 128 & $9.1{\times}10^{7}$ \\
		sin & 25 & 7 & $1.4{\times}10^{7}$ & 6.2 & 38.9 & 13 & $9.7{\times}10^{6}$ & 20.3 & 23.8 & 25 & $6.4{\times}10^{6}$ \\
		sqrt & 64 & 11 & $5.6{\times}10^{7}$ & 2.9 & 41.9 & 32 & $3.3{\times}10^{7}$ & 39.7 & 42.4 & 64 & $2.9{\times}10^{7}$ \\
		square & 128 & 16 & $1.2{\times}10^{8}$ & 2.5 & 47.5 & 64 & $7.2{\times}10^{7}$ & 40.3 & 41.2 & 128 & $6.1{\times}10^{7}$ \\
		\midrule
		c432 & 7 & 3 & n/a & n/a & n/a & 4 & $5.5{\times}10^{4}$ & 0.0 & 11.3 & 7 & $3.5{\times}10^{4}$ \\
		c499 & 32 & 8 & $1.2{\times}10^{5}$ & 9.9 & 43.3 & 16 & $8.8{\times}10^{4}$ & 18.3 & 21.6 & 32 & $5.6{\times}10^{4}$ \\
		c880 & 26 & 7 & $9.0{\times}10^{4}$ & 0.3 & 28.8 & 13 & $5.5{\times}10^{4}$ & 38.4 & 47.0 & 26 & $5.2{\times}10^{4}$ \\
		c1355 & 32 & 8 & $1.2{\times}10^{6}$ & 10.2 & 43.5 & 16 & $8.3{\times}10^{5}$ & 19.2 & 21.8 & 32 & $5.3{\times}10^{5}$ \\
		c1908 & 25 & 7 & $7.1{\times}10^{5}$ & 6.0 & 37.7 & 13 & $4.9{\times}10^{5}$ & 21.2 & 25.6 & 25 & $3.3{\times}10^{5}$ \\
		c2670 & 140 & 17 & $3.3{\times}10^{5}$ & -0.2 & 43.0 & 70 & $2.0{\times}10^{5}$ & 36.9 & 49.0 & 140 & $2.0{\times}10^{5}$ \\
		c3540 & 22 & 6 & $9.6{\times}10^{5}$ & 7.1 & 40.0 & 11 & $6.1{\times}10^{5}$ & 28.3 & 32.9 & 22 & $4.9{\times}10^{5}$ \\
		c5315 & 123 & 16 & $1.2{\times}10^{6}$ & 0.0 & 44.7 & 62 & $6.8{\times}10^{5}$ & 41.7 & 45.4 & 123 & $6.2{\times}10^{5}$ \\
		c6288 & 32 & 8 & $4.4{\times}10^{6}$ & 1.8 & 49.9 & 16 & $2.9{\times}10^{6}$ & 34.1 & 38.1 & 32 & $2.3{\times}10^{6}$ \\
		c7552 & 108 & 15 & $1.4{\times}10^{6}$ & -0.0 & 48.1 & 54 & $7.1{\times}10^{5}$ & 47.1 & 49.7 & 108 & $7.1{\times}10^{5}$ \\
		\bottomrule
	\end{tabular}
	\par\vspace{3pt}
	{\raggedright\footnotesize n/a: too few ancillas to hold the $m$ constraint functions. A tree on $k$ ancillas holds at most $2^{k-1}$ of them, and here $2^{k-1}<m$.\par}
	{\raggedright\footnotesize $^{\dagger}$\,At $k=m$ the budget is saturated and placement is trivial: ShallowGrow, FORM-B, and the W-cycle (then at W-L1) all return the flat tree (one function per ancilla, the stack construction of~\cite{li2024resource}), so the reductions vanish identically and the two columns are omitted.\par}
	{\raggedright\footnotesize On these four circuits (c880, c5315, c2670, c7552 at $k\approx\sqrt{2m}$) most of the cost lies in a few high-cost functions and the small functions contribute little, so the two constructions have similar leaf-node costs; the small depth difference comes from the non-leaf MCX decomposition cost (Section~\ref{sec4.3}).\par}
\end{table*}
This experiment evaluates the effect of the construction choice on the circuit depth of the oracle under equal ancilla budgets, on the EPFL and ISCAS85 circuits. Three constructions are compared: ShallowGrow, FORM-B, and the W-cycle. The W-cycle levels differ in capacity and circuit depth: each higher level holds more functions at greater node depth, and W-L1 reduces to the flat tree, the stack construction of~\cite{li2024resource}, with one function per ancilla merged in one step. All of FORM-B's machinery comes from the FORM model: the recursive tree construction, the validity constraints, and the analytic circuit-depth recursion of Theorem~1. Balance is the simplest choice on the model's one free dimension, the number of functions each subtree carries.

For each circuit, the oracle over its $m$ constraint functions is built with each method in turn. Each circuit is tested at three ancilla budgets, covering the axis from scarce to saturation: $k\approx\sqrt{2m}$, $\lceil m/2\rceil$ (half saturation), and $m$ (saturation). The W-cycle uses the smallest level that holds all $m$ functions; with $k$ ancillas, W-L1 holds at most $k$ functions, W-L2 at most $1+\tbinom{k}{2}$, and W-L3 at most $k+\tbinom{k}{3}$. The three budgets place the W-cycle at W-L3, W-L2, and W-L1, respectively. At the first, neither W-L1 nor W-L2 holds the $m$ functions ($k<m$ and $1+\tbinom{k}{2}<m$) and only W-L3 does. At the second, W-L2 holds them and W-L1 does not ($1+\tbinom{k}{2}\ge m$, $k<m$). At the third, $k=m$ and W-L1 holds them exactly. Table~\ref{tab:comp} reports the ShallowGrow circuit depth and its reduction against FORM-B and the W-cycle for each circuit and budget.

At each budget the W-cycle is taken at its smallest feasible level, which is also its lowest attainable circuit depth there. Against this setting, ShallowGrow is strictly lower in all 39 unsaturated configurations ($k<m$). The average reduction is 43.7\% at the scarce budget and 33.1\% at the middle one; the largest is 49.9\% (c6288). The remaining 20 configurations tie at $k=m$: at this saturated budget every construction returns the same flat tree, one function per ancilla. The reduction over the W-cycle decomposes into two parts: the contribution of the FORM tree itself and that of the ShallowGrow algorithm. FORM-B is 39.5\% below W-L3 at the scarce budget and 5.1\% below W-L2 at the middle one. Against FORM-B, ShallowGrow is on average 29.5\% lower at the middle budget, at most 47.1\% (c7552), and 6.6\% lower at the scarce budget.

The two contributions concentrate at different budgets. At the scarce budget the reduction is due mainly to the FORM tree itself. The W-cycle adapts to the budget only by switching levels: within a level, its circuit depth does not improve with $k$ (Fig.~\ref{fig:tradeoff}), and a tighter budget forces it onto a deeper level. FORM trees instead adjust their node depths gradually, so the gap is widest at the tightest budgets. At the middle budget the reduction is due mainly to the ShallowGrow algorithm. Each non-leaf node computes and later uncomputes its children, so a function placed at node depth $d$ is executed $2^{d}$ times. ShallowGrow always expands the shallowest node first (the minimum-depth strategy), so as many functions as possible end up at shallow leaf nodes; FORM-B distributes the functions evenly over the subtrees, and most end up deeper. At that budget ShallowGrow holds nearly half of the functions at node depth one, FORM-B only two; at the scarce budget both trees hold only two functions at node depth one, and the minimum-depth strategy has little effect.

At the scarce budget the reduction against FORM-B is close to zero for c880 and c5315 and negative for c2670 and c7552, the only negative entries of the table. Since node depth one holds only two functions and the costliest ones are assigned to the shallowest leaves, both trees put the same two large functions there; ShallowGrow can only change the node depths of the remaining, smaller functions. Placing a function one level shallower halves its executions, and the circuit depth saved is proportional to the cost of that function; since only the smaller functions can be moved, the total saving is bounded by their cost share. In these four circuits the cost is concentrated in a few large functions: the cheaper half of the functions accounts for only 0.1\%--7.6\% of the total, against 13\%--50\% for the other circuits. The movable functions carry almost no cost, so the saving vanishes; what remains of the circuit-depth difference is the second term of Theorem~1, the cost of the MCX gates that merge the results, on which ShallowGrow is slightly higher---hence the two negative entries of at most 0.2\%. At the middle budget the opposite holds: ShallowGrow places nearly half of the functions at node depth one, including all the large ones, so the functions moved shallower now carry most of the cost, and c7552 and c5315 show the two largest reductions of the table, 47.1\% and 41.7\%. At the middle budget of c432 ($k=4$), ShallowGrow and FORM-B construct the same tree and the circuit depths coincide.

As the budget grows the W-cycle switches to lower levels, reaching W-L1 at $k=m$, where it coincides with ShallowGrow at the flat tree. With the level fixed, the gap widens with the budget: the circuit depth of a fixed level does not change with $k$, the ShallowGrow circuit depth decreases monotonically in $k$, and the gap peaks at $k=m$. There ShallowGrow is the flat tree and each function is executed twice; a function of W-L2 or W-L3 is executed four or eight times, and with each level held at its best register the realized reductions are 43.7--50.0\% and 59.2--74.2\% (Fig.~\ref{fig:tradeoff}).

Table~\ref{tab:legacy} repeats the comparison on the Boolean quadratic equation systems of \cite{li2024resource}. The oracles are built as actual Qiskit circuits; circuit depth is counted by Qiskit on the circuits after gate decomposition, includes cross-qubit parallelism, and is averaged over the test instances of each size, as in the original comparison. The circuit depths of the remaining experiments are the analytic values of the recursion of Theorem~1. The nine ($n$, ancilla) settings and the three W-cycle levels are those of \cite{li2024resource}; each budget is compared against every level feasible there, twenty (budget, level) combinations in all. The average reduction over the twenty combinations is 54.1\%; the two where W-L1 is feasible are near ties (0.66\%), where both constructions give the same flat tree.

\begin{table}[!htb]
	\centering
	\caption{Comparison on the Boolean quadratic equation instances of \cite{li2024resource}: the circuit depth of the W-cycle at levels one to three and of ShallowGrow, and the reduction (\%) of ShallowGrow relative to each level. n/a: the level does not hold the $m$ functions at this budget.}
	\label{tab:legacy}
	\footnotesize
	\setlength{\tabcolsep}{3.5pt}
	\renewcommand{\arraystretch}{1.1}
	\begin{tabular}{@{}cccrrrrrrr@{}}
		\toprule
		 & & & \multicolumn{4}{c}{Circuit depth} & \multicolumn{3}{c}{Reduction (\%)} \\
		\cmidrule(lr){4-7}\cmidrule(l){8-10}
		$n$ & $m$ & $k$ & W-L1 & W-L2 & W-L3 & ShallowGrow & vs L1 & vs L2 & vs L3 \\
		\midrule
		\multirow{3}{*}{15} & \multirow{3}{*}{5} & 4 & n/a & 540.6 & 1273.3 & 396.6 & n/a & 26.64 & 68.85 \\
		 & & 5 & 151.2 & 540.6 & 1273.3 & 150.2 & 0.66 & 72.22 & 88.20 \\
		 & & 6 & 151.2 & 540.6 & 1273.3 & 150.2 & 0.66 & 72.22 & 88.20 \\
		\midrule
		\multirow{3}{*}{20} & \multirow{3}{*}{7} & 4 & n/a & 1359.1 & 2452.3 & 1132.6 & n/a & 16.67 & 53.81 \\
		 & & 5 & n/a & 1359.1 & 2452.3 & 691.0 & n/a & 49.16 & 71.82 \\
		 & & 6 & n/a & 1359.1 & 2452.3 & 677.8 & n/a & 50.13 & 72.36 \\
		\midrule
		\multirow{3}{*}{25} & \multirow{3}{*}{9} & 5 & n/a & 2094.1 & 5450.3 & 1704.6 & n/a & 18.60 & 68.72 \\
		 & & 6 & n/a & 2094.1 & 5450.3 & 1045.2 & n/a & 50.09 & 80.82 \\
		 & & 7 & n/a & 2094.1 & 5450.3 & 1036.4 & n/a & 50.51 & 80.98 \\
		\bottomrule
	\end{tabular}
\end{table}

\subsection{Space-Depth Trade-off}\label{sec:exp-tradeoff}
\begin{figure*}[!htb]
	\centering
	\includegraphics[width=0.9\textwidth]{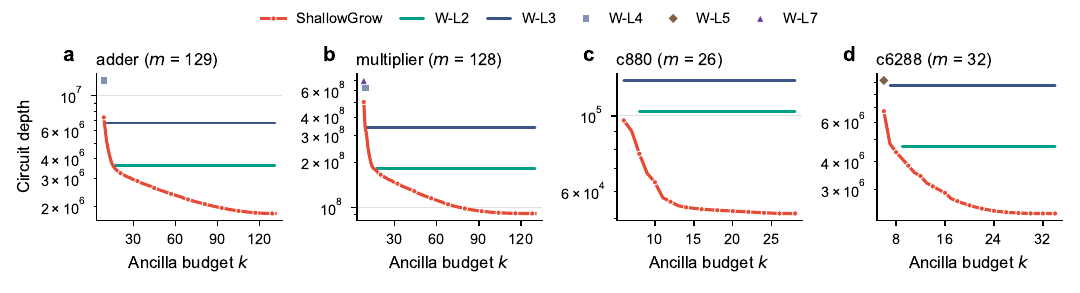}
	\caption{Circuit depth versus the ancilla budget $k$ on four representative circuits (log scale). W-L2 and W-L3 appear from their smallest feasible budgets onward; W-L1 is omitted, as it exists only at the saturation point $k\ge m$, where all methods coincide (Table~\ref{tab:comp}). Below the smallest W-L3 budget, markers show the W-cycle at the smallest feasible level (W-L4, W-L5, or W-L7) at each remaining budget.}
	\label{fig:tradeoff}
\end{figure*}

This experiment examines the full response of circuit depth to the ancilla budget. For each circuit the budget $k$ takes every integer from the smallest feasible budget, the smallest $k$ with $2^{k-1}\ge m$, to $m+2$, and the circuit depth of ShallowGrow and of every feasible W-cycle level is computed at each $k$. The test set, the leaf layer, and the depth accounting are those of Section~\ref{sec:exp-comp}. Figure~\ref{fig:tradeoff} shows the circuit-depth-versus-budget curves of four representative circuits, adder, multiplier, c880, and c6288 (log scale); the remaining sixteen behave alike.

The ShallowGrow circuit depth decreases monotonically in $k$ below saturation. The two ends of the curve are two extreme trees: at the smallest feasible budget the capacity $2^{k-1}$ holds the $m$ constraint functions with no slack and the leaves are pushed to their greatest node depths; at $k=m$ the tree is flat, every function lies at node depth one and is executed twice, the minimum number of executions. Between the two, each added ancilla moves functions onto shallower leaves, and every level gained halves their executions; as the budget approaches saturation, the functions still able to move carry less and less cost, and the curve flattens.

In the same plot the W-cycle levels W-L2 and W-L3 are horizontal segments: once a level is feasible, its circuit depth does not change with $k$: on multiplier, W-L2 remains at $1.82\times10^{8}$ and W-L3 at $3.42\times10^{8}$. Additional ancillas widen the W-cycle nodes without shortening their node depths, so the best register remains the one at the level's smallest feasible budget and the extra ancillas remain idle; when the budget tightens, the only response is a switch to a deeper level, each level has its own smallest feasible budget, and W-L1 exists only at $k\ge m$. Below the smallest W-L3 budget the ladder continues through deeper levels, marked in the plot at their smallest feasible budgets: W-L4 at $k=9$ on adder and multiplier, W-L5 at $k=6$ on c6288, and W-L7 at $k=8$ on multiplier, at $1.4\times$--$1.8\times$ the ShallowGrow circuit depth of the same budget. At every fixed budget the ShallowGrow circuit depth lies below every feasible W-cycle level, meeting W-L1 only at $k=m$; as the budget varies, below $k=m$ only the ShallowGrow circuit depth follows it.

\subsection{Complete Oracle Construction with Leaf-Level Synthesis}\label{sec:exp-stack}
LHRS \cite{soeken2019lut} and the divide-and-conquer pebbling strategy of Zhang et al.\ \cite{Zhang2025A} decompose and schedule the computation within the logic network of a given function; the FORM model and ShallowGrow treat each function as an atomic unit and optimize the structure that composes the evaluation results of multiple functions under a fixed ancilla budget. The two approaches are complementary, and a complete oracle can be constructed in two steps: leaf-level synthesis first produces the circuit of each constraint function, and the composition layer then organizes these circuits into the oracle. This experiment evaluates the resource consumption of the complete oracle constructions. The constructions considered fall into two classes: the composite class pairs a composition layer (ShallowGrow or the W-cycle) with a leaf layer (Zhang et al.\ or LHRS); the leaf-only class is the whole-circuit construction that Zhang et al.\ and LHRS publish, which encodes all $m$ outputs at once.

This section reports T-count and total qubit count: in fault-tolerant implementations the T gates dominate the time cost, and the total qubit count is the space cost. Under these metrics the published values of Zhang et al.\ and LHRS (Table~\ref{tab:leafpub}) are used directly. The ISCAS85 circuits do not appear in this section, because neither leaf-level study reports T-counts for them; hyp, which lacks a published Zhang result, is evaluated on the LHRS leaf only.

\begin{figure*}[!htb]
	\centering
	\includegraphics[width=0.85\textwidth]{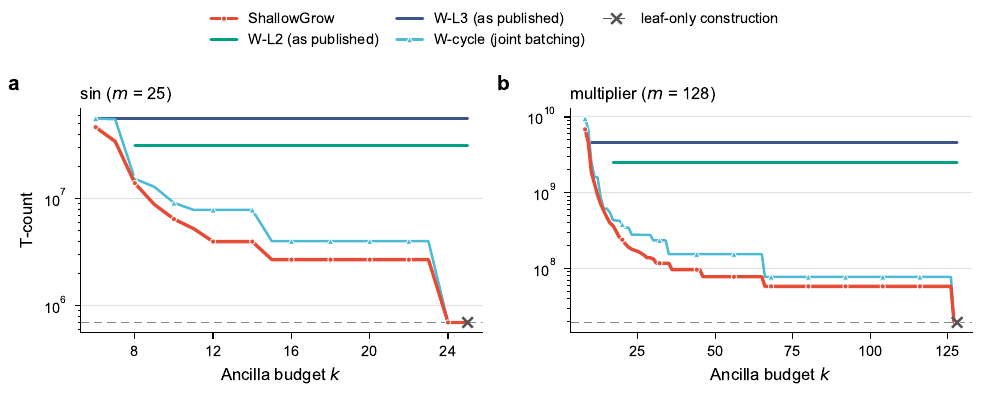}
	\caption{T-count on the Zhang leaf versus the ancilla budget $k$ on sin and multiplier (log scale). Thick horizontal lines: the W-cycle as published, one function per W-cycle node. Light-blue curve: the W-cycle with our joint batching applied, at its best feasible level (W-L2--W-L7) per budget. The leaf-only construction is a single point at $k=m-1$ (cross); the dashed line extends its T-count leftward for reference.}
	\label{fig:stack}
\end{figure*}

The T-count of a leaf-only construction is $2T_{\mathrm{pub}}+\mathrm{MCZ}(m)$ (one computation and one uncomputation within the oracle) and its total qubit count is the published value; a composite construction occupies $n+k+W+1$ qubits, where the leaf workspace $W$ (the published qubits minus inputs and outputs) is reused serially across leaves. The T-count of a composite construction is obtained by evaluating the recursion of Theorem~1 with T weights: a non-leaf node computes and uncomputes each child once and merges the results by an MCX; a Toffoli contributes 7 T gates and an MCX with $\kappa\ge3$ controls contributes $28(\kappa-2)$; leaves are priced per AND node by the calibration $\rho$ of Section~\ref{sec:exp-setup}. The experiments consist of three parts: Table~\ref{tab:tight} lowers the budget stepwise (four representative circuits), Table~\ref{tab:stack} presses the budget to the ShallowGrow floor (all ten EPFL circuits), and Fig.~\ref{fig:stack} sweeps the full budget axis (sin and multiplier).

\begin{table*}[!htb]
	\centering
	\caption{Oracle constructions on four representative EPFL circuits under a decreasing total-qubit budget. Each row fixes one budget (Qubits) and evaluates every construction at it: the published leaf-only construction and, on each published leaf, the W-cycle at its best feasible recursion level (W-cyc.; the level in parentheses) and ShallowGrow. Both W-cycle and ShallowGrow entries take the best ancilla register $k$ fitting $n+k+W+1$ within the budget ($k$ column; $W$: leaf workspace $Q_{\mathrm{pub}}-n-m$, serially reused) and evaluate the outputs in joint batches (batch width chosen to minimize T-count; at $k=m-1$ a single batch covers the whole network and every construction coincides with the leaf-only construction). T-counts in units of $10^{6}$; n/f: does not fit the budget; boldface: strictly lowest T-count in the row.}
	\label{tab:tight}
	\footnotesize
	\setlength{\tabcolsep}{4.5pt}
	\begin{tabular}{lrrrrrrrrrrr}
		\toprule
		 & & \multicolumn{5}{c}{Leaf layer: Zhang et al.\ \cite{Zhang2025A}} & \multicolumn{5}{c}{Leaf layer: LHRS \cite{soeken2019lut}} \\
		\cmidrule(lr){3-7}\cmidrule(lr){8-12}
		Circuit & Qubits & $k$ & $W$ & Leaf-only & W-cyc. & ShallowGrow & $k$ & $W$ & Leaf-only & W-cyc. & ShallowGrow \\
		\midrule
		adder ($m{=}129$) & 505 & 128 & 15 & 0.070 & 0.070 (L2) & 0.070 & 128 & 120 & 0.0075 & 0.0075 (L2) & 0.0075 \\
		 & 400 & 128 & & 0.070 & 0.070 (L2) & 0.070 & 23 & & n/f & 0.048 (L2) & \textbf{0.032} \\
		 & 337 & 65 & & n/f & 0.399 (L2) & \textbf{0.206} & -- & & n/f & n/f & n/f \\
		 & 305 & 33 & & n/f & 0.546 (L2) & \textbf{0.273} & -- & & n/f & n/f & n/f \\
		 & 289 & 17 & & n/f & 0.883 (L2) & \textbf{0.762} & -- & & n/f & n/f & n/f \\
		 & 281 & 9 & & n/f & 15.1 (L4) & \textbf{8.82} & -- & & n/f & n/f & n/f \\
		\addlinespace
		multiplier ($m{=}128$) & 5,806 & 127 & 352 & 19.4 & 19.4 (L2) & 19.4 & 127 & 5,550 & 0.803 & 0.803 (L2) & 0.803 \\
		 & 608 & 127 & & 19.4 & 19.4 (L2) & 19.4 & -- & & n/f & n/f & n/f \\
		 & 545 & 64 & & n/f & 154 (L2) & \textbf{77.5} & -- & & n/f & n/f & n/f \\
		 & 513 & 32 & & n/f & 234 (L2) & \textbf{117} & -- & & n/f & n/f & n/f \\
		 & 497 & 16 & & n/f & 546 (L2) & \textbf{398} & -- & & n/f & n/f & n/f \\
		 & 489 & 8 & & n/f & 9,510 (L7) & \textbf{6,850} & -- & & n/f & n/f & n/f \\
		\addlinespace
		sin ($m{=}25$) & 1,468 & 24 & 334 & 0.698 & 0.698 (L2) & 0.698 & 24 & 1,419 & 0.219 & 0.219 (L2) & 0.219 \\
		 & 383 & 24 & & 0.698 & 0.698 (L2) & 0.698 & -- & & n/f & n/f & n/f \\
		 & 372 & 13 & & n/f & 7.84 (L2) & \textbf{3.96} & -- & & n/f & n/f & n/f \\
		 & 366 & 7 & & n/f & 54.9 (L3) & \textbf{34.1} & -- & & n/f & n/f & n/f \\
		 & 365 & 6 & & n/f & 55.8 (L3) & \textbf{46.3} & -- & & n/f & n/f & n/f \\
		\addlinespace
		square ($m{=}128$) & 4,058 & 127 & 162 & 4.15 & 4.15 (L2) & 4.15 & 127 & 3,866 & 1.01 & 1.01 (L2) & 1.01 \\
		 & 354 & 127 & & 4.15 & 4.15 (L2) & 4.15 & -- & & n/f & n/f & n/f \\
		 & 291 & 64 & & n/f & 32.4 (L2) & \textbf{16.3} & -- & & n/f & n/f & n/f \\
		 & 259 & 32 & & n/f & 49.1 (L2) & \textbf{24.6} & -- & & n/f & n/f & n/f \\
		 & 243 & 16 & & n/f & 114 (L2) & \textbf{82.7} & -- & & n/f & n/f & n/f \\
		 & 235 & 8 & & n/f & 1,920 (L7) & \textbf{1,390} & -- & & n/f & n/f & n/f \\
		\bottomrule
	\end{tabular}
\end{table*}

Table~\ref{tab:tight} lowers the total-qubit budget stepwise on four representative EPFL circuits: the first row sets the budget to the published total qubit count of the leaf-level study's leaf-only construction ($k=m-1$), each following row halves the ShallowGrow ancilla register $k$ down to the smallest feasible budget, and every row reports the T-count (in $10^{6}$) and feasibility of each construction (n/f: not feasible); the remaining circuits behave alike. In the first row the total is exactly the published $n+m+W$, and there the leaf-only construction has the lowest T-count. When the budget is below the published value, neither the Zhang strategy nor the LHRS framework admits a feasible construction (n/f in the table): the total qubit count of each is fixed by its pebbling strategy, and the published value is its lower bound.

ShallowGrow compresses the result register through the FORM tree, and as the budget tightens it continues to yield constructions with a deeper tree. On adder the W-cycle stays at level two down to 289 qubits ($k=17$). The W-cycle hierarchy does not stop at level three: level $\ell$ on $k$ ancillas holds up to $\sum_{j=0}^{\ell}\binom{k-1}{j}$ constraint functions, saturating at the same $2^{k-1}$ capacity as the FORM tree \cite{li2024resource}. In the last row of adder, 281 qubits ($k=9$), the smallest feasible level is W-L4; in the last rows of multiplier and square ($k=8$, $2^{k-1}=m$) it is W-L7, and the resulting T-counts lie $1.4\times$--$1.7\times$ above the ShallowGrow value in the same row (Table~\ref{tab:tight}). In every feasible row the ShallowGrow T-count is the lowest (boldface in the table).

The last row saves 20\% to 34\% of the total qubits relative to the published Zhang value (4.7\% on sin), compressing the result register from $m$ to $k+1$ qubits, by 72\% to 93\%. The saving is bounded by the share of the result register in the published total: the composition layer leaves the input register and the leaf workspace unchanged, and on sin the workspace accounts for 87\% of the published qubits, so at most 4.7\% can be saved. The saved qubits are paid for in T-count: on adder, multiplier, and square each step multiplies it by 1.3 to 4, and the last step to the smallest feasible budget by 11.6 to 17; on sin, whose floor lies only one ancilla below the previous row, the steepest step comes earlier. The space-depth trade-off of Section~\ref{sec:exp-tradeoff} thus appears again on the layer above leaf-level synthesis. The 400-qubit row of adder yields a further cross-leaf observation: the published LHRS construction (505 qubits) exceeds this budget, while ShallowGrow+LHRS is feasible at $k=23$ with a T-count of $0.032\times10^{6}$, below the leaf-only Zhang construction at the same budget ($0.070\times10^{6}$): the composition layer makes a leaf of lower T-count feasible below its published total qubit count.

\begin{table*}[!htb]
	\centering
	\caption{Oracle constructions on all ten EPFL circuits at two shared total-qubit budgets per circuit, evaluated as in Table~\ref{tab:tight}. The upper budget is the total qubit count of the published leaf-only LHRS construction: a single batch is feasible and every construction coincides with the leaf-only construction of its leaf, so no entry is bolded. The lower budget is the ShallowGrow feasibility floor on the Zhang leaf (for hyp, the LHRS leaf): the smallest $k$ with $2^{k-1}\ge m$, at total $n+k+W+1$. T-counts in units of $10^{6}$; boldface: strictly lowest T-count in the row; n/f: does not fit the budget; --: no published leaf data.}
	\label{tab:stack}
	\footnotesize
	\begin{tabular}{lrrrrrrrrr}
		\toprule
		 & & \multicolumn{4}{c}{Leaf layer: Zhang et al.\ \cite{Zhang2025A}} & \multicolumn{4}{c}{Leaf layer: LHRS \cite{soeken2019lut}} \\
		\cmidrule(lr){3-6}\cmidrule(lr){7-10}
		Circuit & Qubits & $k$ & Leaf-only & W-cyc. & ShallowGrow & $k$ & Leaf-only & W-cyc. & ShallowGrow \\
		\midrule
		adder ($m{=}129$) & 505 & 128 & 0.070 & 0.070 (L2) & 0.070 & 128 & 0.0075 & 0.0075 (L2) & 0.0075 \\
		 & 281 & 9 & n/f & 15.1 (L4) & \textbf{8.82} & -- & n/f & n/f & n/f \\
		\addlinespace
		bar ($m{=}128$) & 840 & 127 & 0.160 & 0.160 (L2) & 0.160 & 127 & 0.038 & 0.038 (L2) & 0.038 \\
		 & 206 & 8 & n/f & 43.1 (L7) & \textbf{29.6} & -- & n/f & n/f & n/f \\
		\addlinespace
		div ($m{=}128$) & 3,399 & 127 & 13.9 & 13.9 (L2) & 13.9 & 127 & 0.815 & 0.815 (L2) & 0.815 \\
		 & 1,494 & 8 & n/f & 5,400 (L7) & \textbf{3,830} & -- & n/f & n/f & n/f \\
		\addlinespace
		hyp ($m{=}128$) & 47,814 & -- & -- & -- & -- & 127 & 6.05 & 6.05 (L2) & 6.05 \\
		 & 47,695 & -- & -- & -- & -- & 8 & n/f & 8,980 (L7) & \textbf{6,190} \\
		\addlinespace
		log2 ($m{=}32$) & 7,611 & 31 & 6.62 & 6.62 (L2) & 6.62 & 31 & 1.26 & 1.26 (L2) & 1.26 \\
		 & 1,187 & 6 & n/f & 1,360 (L5) & \textbf{968} & -- & n/f & n/f & n/f \\
		\addlinespace
		max ($m{=}130$) & 1,036 & 129 & 1.07 & 1.07 (L2) & 1.07 & 129 & 0.040 & 0.040 (L2) & 0.040 \\
		 & 718 & 9 & n/f & 888 (L4) & \textbf{627} & -- & n/f & n/f & n/f \\
		\addlinespace
		multiplier ($m{=}128$) & 5,806 & 127 & 19.4 & 19.4 (L2) & 19.4 & 127 & 0.803 & 0.803 (L2) & 0.803 \\
		 & 489 & 8 & n/f & 9,510 (L7) & \textbf{6,850} & -- & n/f & n/f & n/f \\
		\addlinespace
		sin ($m{=}25$) & 1,468 & 24 & 0.698 & 0.698 (L2) & 0.698 & 24 & 0.219 & 0.219 (L2) & 0.219 \\
		 & 365 & 6 & n/f & 55.8 (L3) & \textbf{46.3} & -- & n/f & n/f & n/f \\
		\addlinespace
		sqrt ($m{=}64$) & 3,204 & 63 & 6.26 & 6.26 (L2) & 6.26 & 63 & 0.816 & 0.816 (L2) & 0.816 \\
		 & 786 & 7 & n/f & 666 (L6) & \textbf{486} & -- & n/f & n/f & n/f \\
		\addlinespace
		square ($m{=}128$) & 4,058 & 127 & 4.15 & 4.15 (L2) & 4.15 & 127 & 1.01 & 1.01 (L2) & 1.01 \\
		 & 235 & 8 & n/f & 1,920 (L7) & \textbf{1,390} & -- & n/f & n/f & n/f \\
		\bottomrule
	\end{tabular}
\end{table*}

\begin{table}[!htb]
	\centering
	\caption{Published results of the leaf-only constructions of the two leaf-level studies on the EPFL circuits, as reported in Table~III of \cite{soeken2019lut} and Table~II of \cite{Zhang2025A}. These values calibrate the leaf models of Tables~\ref{tab:tight} and~\ref{tab:stack}. ``--'': not reported.}
	\label{tab:leafpub}
	\footnotesize
	\setlength{\tabcolsep}{4.5pt}
	\begin{tabular}{lrrrr}
		\toprule
		 & \multicolumn{2}{c}{LHRS \cite{soeken2019lut}} & \multicolumn{2}{c}{Zhang et al.\ \cite{Zhang2025A}} \\
		\cmidrule(lr){2-3}\cmidrule(lr){4-5}
		Circuit & Qubits & T-count & Qubits & T-count \\
		\midrule
		adder & 505 & 1,988 & 400 & 33,271 \\
		bar & 840 & 17,024 & 325 & 78,208 \\
		div & 3,399 & 405,896 & 1,613 & 6,969,849 \\
		hyp & 47,814 & 3,021,920 & -- & -- \\
		log2 & 7,611 & 628,777 & 1,212 & 3,309,687 \\
		max & 1,036 & 18,369 & 838 & 534,148 \\
		multiplier & 5,806 & 399,632 & 608 & 9,684,013 \\
		sin & 1,468 & 109,248 & 383 & 348,731 \\
		sqrt & 3,204 & 407,081 & 842 & 3,130,481 \\
		square & 4,058 & 504,369 & 354 & 2,073,754 \\
		\bottomrule
	\end{tabular}
\end{table}

Table~\ref{tab:stack} presses all ten circuits to the ShallowGrow feasibility floor, the smallest $k$ with $2^{k-1}\ge m$: no leaf-only construction fits these budgets, the W-cycle enters at its smallest feasible level, L3 to L7, and in every row ShallowGrow has the lowest T-count, 17--42\% below the W-cycle. Relative to the published totals the floor budgets save 0.2\%--37\% of the qubits, the bound set by the result-register share of each published total; the reductions are independent of the leaf, since the calibration rescales the evaluation costs uniformly and the ratios are set by the tree structure. The two layers act on different factors of the cost expression of Theorem~1: the leaf synthesizer rescales $\delta$ and sets the workspace $W$, while the composition layer sets the re-computation profile $\sum 2^{d(v)}$ and the ancilla register $k$; within a construction the two factors multiply, so the composition-layer differences do not depend on the leaf choice. Switching the leaf is an independent variable: replacing the Zhang leaf by LHRS trades a larger workspace for a lower T-count.

Figure~\ref{fig:stack} plots the T-count against the budget $k$ on sin and multiplier. The original W-cycle assigns one function per W-cycle node and its T-count does not respond to the budget; at $k=\lceil m/2\rceil$ the W-L2 T-count is $7.9\times$ ShallowGrow's on sin and $32\times$ on multiplier. This experiment introduces a joint-batching optimization for the composite constructions: adjacent constraint functions are merged into a batch, and each cone shared within the batch is evaluated once per pass. With the same optimization applied to the W-cycle, taken at its best feasible level per budget (up to W-L7 at the tightest budgets), its curve falls with the budget and reaches the same feasibility floor as ShallowGrow, and the gap to ShallowGrow narrows to within $2\times$; at every budget ShallowGrow has the lowest T-count. The leaf-only construction appears in the figure as a single point at $k=m-1$ (the cross): there one batch covers the whole network and every batched construction coincides with it.

The curves of Fig.~\ref{fig:stack} form a staircase: the T-count falls at certain budgets and is unchanged over the intervals between them. A construction on the curve is fixed by the batch width $\beta$ and the tree's ancilla count $k_2$; it occupies $\beta+k_2$ ancillas ($\beta=1$ reserves no output qubits), and its circuit and T-count are thereby determined. The large T-count drops arise from batch merging, and a merge requires a fixed number of qubits: decreasing the batch count from $B=\lceil m/\beta\rceil$ to $B-1$ requires each batch to hold about $m/B(B-1)$ additional outputs, each occupying one qubit; after a merge, cones shared within a batch are evaluated once, and the fewer the batches, the less repeated evaluation. While the budget increment is below that number, the construction and its T-count are unchanged; at that number the merge occurs and the T-count falls.

Batch count $B$ becomes feasible at the threshold $k(B)=\lceil m/B\rceil+\lceil\log_2 B\rceil+1$ ($B=1$ is the single batch, $k=m-1$), and the realized step lags this threshold by at most one or two ancillas (a new $B$ becomes feasible first with its deepest tree and occasionally requires one further flattening of the tree to reach a lower T-count). The step positions therefore follow approximately the spacing $m/B(B+1)$: dense at small budgets, with the unchanged intervals growing as $m/B^{2}$ at large ones; as $B$ decreases the tree retains fewer leaves and its ancillas are reassigned to batch outputs, until at $B=1$ the tree vanishes and the curve coincides with the leaf-only construction. Verified on sin: the thresholds $8,9,12,15$ for $B=7,5,3,2$ coincide with the steps of an exhaustive search, $B=4$ lags by one ancilla; the two small early steps ($k=7,10$) correspond to a constant batch count with the tree flattened by one level.

\section{Conclusion}
This paper introduces the Framework for Oracle Recursion Modeling (FORM), specifying its structural definition, attribute correspondences, and the constraints required for model validity. The framework provides a comprehensive representation of XOR-oracle recursive structures, enabling an intuitive characterization of complex recursive relationships. Building on the FORM model, we propose a quantum gate complexity analysis approach that quantifies the number of quantum gates in the oracle circuit by aggregating the costs associated with nodes and edges in the model. This approach reveals the primary sources of quantum gate complexity and provides an analytical foundation for subsequent optimization algorithms. The FORM model is applicable not only to the ShallowGrow algorithm but also serves as a theoretical foundation for future studies on oracle optimization.

Based on the FORM model, we propose the ShallowGrow algorithm, which minimizes the number of function evaluations under a given ancilla budget. Our analysis shows that increasing the number of ancilla qubits reduces the circuit depth of the resulting oracles. We provide a theoretical proof that the FORM tree constructed by the ShallowGrow algorithm minimizes the number of function evaluations, i.e., the leaf-node cost $\sum_{\kappa(v)=0}2^{d(v)}\cdot\delta$. Experimental results on Boolean quadratic equation systems with $n=15,20,25$ variables show that under equal ancilla budgets the Qiskit-measured circuit depth of oracles constructed by our method is 54.1\% below that of the W-cycle approach on average; on the combinational logic networks of the EPFL and ISCAS85 suites, the analytic circuit-depth reduction averages 43.7\% under scarce ancilla budgets. When the leaf circuits of published pebbling-based syntheses are composed by ShallowGrow, a complete oracle runs on logarithmically many ancillary qubits, budgets at which no stand-alone published construction fits; at half as many ancillas as constraint functions, the W-cycle-based construction requires 7.9 to 32 times the T-count of ShallowGrow.

Several open issues and directions for future work remain. The current FORM model and ShallowGrow algorithm do not account for specific function encoding methods or decomposition strategies for multi-controlled X (MCX) gates. The circuit depth after MCX gate decomposition is significantly influenced by the number of ancilla qubits available for the decomposition process. Future work will explore the joint optimization of ancilla qubit allocation between the constraint function computation and MCX gate decomposition stages to achieve further reductions in circuit depth. 
Another important direction is to investigate the asymptotic behavior of the proposed model and algorithm under large-scale parameters, and to establish more general analytical results for complexity characterization.
The proposed framework is highly general and can be applied to a broader class of constraint satisfaction problems (CSPs).
In particular, evaluating the FORM model and ShallowGrow algorithm on more complex problems, such as AES cryptanalysis, may further demonstrate the effectiveness of the proposed framework and optimization strategy.
In addition, the proposed approach is complementary to hierarchical synthesis and pebbling-based methods: the FORM model organizes the composition of multiple functions, whereas the LHRS framework\cite{soeken2019lut} and pebbling strategies, such as the divide-and-conquer strategy of Zhang et al.\cite{Zhang2025A}, optimize the synthesis of each individual function, which corresponds to a leaf node of the FORM tree. Section~\ref{sec:exp-stack} combines the two levels through their published circuits; a tighter co-design, scheduling the pebbling of each function jointly with the tree structure and the batch partition, may further reduce resource consumption in practical oracle implementations.
These directions will be investigated in future studies.

\clearpage
\onecolumn

\columnratio{0.48}
\setlength{\columnsep}{1.2em}

\end{document}